\documentclass[conference]{IEEEtran}
\IEEEoverridecommandlockouts
\usepackage{url}
\usepackage{cite}
\usepackage{amsmath,amssymb,amsfonts}
\usepackage{algorithmic}
\usepackage{graphicx}
\usepackage{textcomp}
\usepackage{xcolor}
\def\BibTeX{{\rm B\kern-.05em{\sc i\kern-.025em b}\kern-.08em
    T\kern-.1667em\lower.7ex\hbox{E}\kern-.125emX}}

\usepackage{lipsum}

\graphicspath{{figures/}}
\usepackage{multirow}

\makeatletter
\def\input@path{{./tables/}}
\makeatother

\makeatletter
\def\ps@IEEEtitlepagestyle{%
  \def\@oddfoot{\mycopyrightnotice}%
  \def\@evenfoot{}%
}
\def\mycopyrightnotice{%
  {\footnotesize Accepted for presentation at PMAPS 2020. \textcopyright 2020 IEEE\hfill}
  \gdef\mycopyrightnotice{}
}
\makeatother

\usepackage{booktabs}
\usepackage[caption=false]{subfig}

\begin{document}

\title{Calculations of System Adequacy Considering \\Heat Transition Pathways\\
\thanks{This work was funded by the Engineering and Physical Sciences Research Council through grant no.  EP/S00078X/1 (Supergen Energy Networks hub 2018). S. Sheehy is funded by an EPSRC studentship.}
}

\author{
\IEEEauthorblockN{
Matthew Deakin, Sarah Sheehy\IEEEauthorrefmark{1}, David M. Greenwood, Sara Walker and Phil C. Taylor
}
\IEEEauthorblockA{
School of Engineering\\
Newcastle University\\ Newcastle, UK\\
\{matthew.deakin, david.greenwood, sara.walker, phil.taylor\}@newcastle.ac.uk
}
\IEEEauthorblockA{
\IEEEauthorrefmark{1}Department of Engineering\\
Durham University\\ Durham, UK\\
sarah.sheehy2@durham.ac.uk
}
}

\maketitle

\begin{abstract}
The decarbonisation of heat in developed economies represents a significant challenge, with increased penetration of electrical heating technologies potentially leading to unprecedented increases in peak electricity demand. This work considers a method to evaluate the impact of rapid electrification of heat by utilising historic gas demand data. The work is intended to provide a data-driven complement to popular generative heat demand models, with a particular aim of informing regulators and actors in capacity markets as to how policy changes could impact on medium-term system adequacy metrics (up to five years ahead). Results from a GB case study show that the representation of heat demand using scaled gas demand profiles increases the rate at which 1-in-20 system peaks grow by 60\%, when compared to the use of scaled electricity demand profiles. Low end-use system efficiency, in terms of aggregate coefficient of performance and demand side response capabilities, are shown to potentially lead to a doubling of electrical demand-temperature sensitivity following five years of heat demand growth.
\end{abstract}

\begin{IEEEkeywords}
System adequacy, multi-vector systems, power system reliability, demand modelling
\end{IEEEkeywords}

\section{Introduction}\label{e:s1}

Systemic changes in electrical demands, led by the electrification of carbon-intensive loads such as heat and transport, represent an enormous challenge for power systems planners across the globe. This, in combination with increased levels of intermittent renewable generation, has led to many governments implementing (or considering) capacity markets to secure long-term investment in deregulated electrical power systems. There has therefore been renewed interest in system adequacy from the point of view of market regulators and from market actors, both of whom need effective tools to make robust decisions on behalf of their stakeholders.

Attempts to electrify heat demand are predicted to lead to significant increases electricity demand in some economies such as the UK \cite{ngeso2019fes}. Works that consider the impact of electrification of heat on power systems generally use a generative (or `bottom-up') approach. This method is characterised by the use of heating demand profiles, typically found using device-level measurements or building-scale simulations, which are then scaled by total demand and added to representative power system profiles \cite{ngeso2019fes,love2017addition,clegg2019integrated,hawker2018spatial}. For example, in \cite{clegg2019integrated} the authors use building simulation software combined with demographic information to estimate spatially disaggregated heating demand across the whole of Great Britain (GB). Other works derive heating demand profiles directly from measured data--in \cite{love2017addition}, the authors aggregate the outputs of 696 domestic heat pumps to determine a characteristic demand profile which is then used to draw conclusions about system peaks. Whilst this type of approach has advantages in terms of the intelligibility of the models, presuppositions about \textit{future} heating demand profiles are required, which, in turn, could fundamentally change given increased connectivity and flexibility of devices within smart local energy systems.

To the contrary, data-driven approaches that make use of historic electrical demand and generation data are common in system adequacy studies. There has been a strong focus in recent works on the inclusion of renewables within system adequacy and valuation \cite{dent2016capacity,keane2010capacity,sheehy2016impact,hasche2010capacity}. To our knowledge, however, there are no works that calculate probabilistic system adequacy measures considering heat demand growth explicitly. The authors of \cite{wilson2013historical} do consider the impact of the electrification of heat on peak daily system demands using historic gas demand data directly, although the work does not go beyond the evaluation of deterministic peak demands.

This paper proposes a novel approach for the study of system adequacy, explicitly considering heat transition pathways, for the purpose of medium-term power system planning. The work is intended to combine the benefits of the disparate generative and data-driven system adequacy approaches previously outlined. Historic gas demand is used as a proxy for growing electrical heating demand, whilst taking into account flexibilities that could reduce heating demands during critical demand peaks via demand side response (or thermal storage). We compare how the use of business-as-usual electrical demand growth profiles compare against the use of historic gas profiles when modelling heating demand, and how this impacts on the vulnerability of systems to cold snaps (in terms of demand-temperature sensitivity).

Section \ref{s:s2} describes two methods by which heat demand could be accounted for in system adequacy calculations. Section \ref{s:s3} describes the data and assumptions used to study the difference between these models, based on a GB system case study; this forms the basis for the results of Section \ref{s:s4}, which compares the two models against a base model. System risk metrics and temperature sensitivity are studied, demonstrating the key differences between the out-turn of the two models. Finally, salient conclusions drawn from this work are discussed in Section \ref{s:s5}.

\section{Including Heat Demand in System Adequacy}\label{s:s2}

In this work, the system margin is modelled using a time-collapsed (or `snapshot') model. The system margin $Z$ is calculated by taking the difference of the total generation and total effective demand; a positive system margin is an adequate system; a system with a negative margin is inadequate. An inadequate system state results in the requirement of an intervention in the normal operation of the market by the electricity system operator (SO), with severe shortfalls eventually resulting in load shedding.

\subsection{Generation}

The generation in a system is often disaggregated into two parts: conventional (dispatchable) generation $X$ and renewable generation $Y$. Conventional generation here covers, amongst others, thermal units and interconnectors, whilst renewable generation $Y$ consists of the output of onshore and offshore wind generators\footnote{Peak loads in the UK are after sunset \cite[pp. 123]{ngeso2019etys} and so solar photovoltaics are not included within adequacy calculations in this work.}. The former is inherently random due to unplanned, forced outages at the unit level, whilst the stochasticity of the latter is driven by the unpredictability of long-term weather forecasts.

The system total generation is therefore given by
\begin{equation}\label{e:generation}
\mathrm{Total\:Generation} = X + Y\,,
\end{equation}
with the cumulative distribution function (CDF) of the total generation denoted by $F_{(X+Y)}$.

Conventional generation $X$ is modelled using the method of, e.g., \cite{hasche2010capacity}, in which each generator is modelled with a capacity and an availability (or forced outage rate); the probability distribution function (PDF) of $X$ therefore obtained by the convolution of the PDFs of the individual generation units. Similarly, a PDF of the renewable generation $Y$ is found by scaling on and offshore wind hourly capacity factors $W_{\mathrm{on}},\,W_{\mathrm{off}}$ (in \%) as
\begin{equation}\label{e:findY}
Y = c_{\mathrm{on}}W_{\mathrm{on}} + c_{\mathrm{off}}W_{\mathrm{off}}\,,
\end{equation}
where $c_{\mathrm{on}},\,c_{\mathrm{off}}$ represent the on and offshore install wind capacities respectively (in MW).

\subsection{Total Effective Demand}\label{s:ted}

In this work the total electrical demand is modelled as the sum of underlying electrical demand $E$, modelled using historic data, and the electrical heating demand $H$. Additionally, a reserve $r$ is required to mitigate against the risk of the loss of largest infeed, resulting in the total effective demand $D$ being given by the linear sum
\begin{equation}\label{e:ted}
D = E + H + r\,.
\end{equation}
The present value of the reserve $r$ is 1320 MW \cite{national2019security}. For simplicity, it is assumed that there is no growth in underlying electrical demand $E$, matching five-year industry forecasts for GB peak and average electrical demands \cite{ngeso2019fes}.

\subsection{Methods for Modelling Heat Demand}

Historic demand, corrected for weather and underlying trends, can be used to model the electrical demand $E$ \cite{sheehy2016impact}. On the other hand, electrical heating is relatively uncommon in some systems (such as the UK), and so a method of estimating the impact of $H$ is required.

To make calculations about impacts of an increase in heating demand, denote the additional annual heating demand $\epsilon_{\Delta H}$, and the annual (nominal) electrical and gas demands as $\epsilon_{E},\,\epsilon_{G}$ respectively. We propose that system adequacy under heat transition pathways be studied in the context of three models.

\subsubsection{Model $\mathcal{M}_{0}$ (the `base case' model)}
Heating demand is ignored, i.e., 
\begin{equation}\label{e:modelM0}
H=0\,.
\end{equation}
This is used to calculate baseline figures, with system adequacy changes only dependent on Total Generation \eqref{e:generation}, and so is referred to as the \textit{base model}.

\subsubsection{Model $\mathcal{M}_{1}$ (the `electricity-scaled' model)}
Heating demand is assumed to largely follow historic electrical demand profiles, such that
\begin{equation}\label{e:modelM1}
H = k_{E\to H} E\,,
\end{equation}
where the electric demand-heat sensitivity $k_{E\to H}$ is given by
\begin{equation}\label{e:d2h_coeffs}
k_{E\to H} = \dfrac{ \epsilon_{\Delta H} }{\epsilon_{E}\,k_{\mathrm{DSR}}\, k_{\mathrm{COP}}}\,,
\end{equation}
where $k_{\mathrm{DSR}},\, k_{\mathrm{COP}}$ models the effect of demand side response (DSR) and coefficient of performance (COP) of the electrified heat system. This modelled is referred to as the \textit{electricity-scaled} model.

\subsubsection{Model $\mathcal{M}_{2}$ (the `gas-scaled' model)}
Heating temporal profiles are assumed to follow gas demand $G$ as a proxy for heating demand,
\begin{equation}\label{e:modelM2}
H = k_{G\to H} G\,,
\end{equation}
with the linear scaling coefficient $k_{G\to H}$ found as
\begin{equation}\label{e:g2h_coeffs}
k_{G\to H} = \dfrac{ \epsilon_{\Delta H} }{\epsilon_{G}\,k_{\mathrm{DSR}}\, k_{\mathrm{COP}}}\,.
\end{equation}
(Note that the right hand sides of \eqref{e:d2h_coeffs}, \eqref{e:g2h_coeffs} are identical except for the energy term $\epsilon_{(\cdot)}$ on the denominator.) It is implicit in \eqref{e:modelM2} that the gas demand $G$, which is only measured daily at the transmission level, is split evenly throughout the day, with the DSR coefficient $k_{\mathrm{DSR}}$ accounting for the size of the demand at the system peak. This final model is referred to as the \textit{gas-scaled} model.

The aggregate coefficient of performance parameter $k_{\mathrm{COP}}$ models the reduction in energy consumption of electric heating devices, which typically have a COP much greater than unity \cite{staffell2012review}. The DSR coefficient $k_{\mathrm{DSR}}$ is used to model the correlation between daily heat demands and existing peaks in electricity demand. The expected electrical load of electric heat pumps is known to be very different to that of gas boilers--domestic heat pumps are typically rated at much lower powers than domestic gas boilers, and so load profiles are very different \cite{wwu2018freedom}. 

A DSR factor greater than unity ($k_{\mathrm{DSR}}>1$) represents a system which effectively reduces heat demands during existing electrical system peaks, whilst a value less than unity ($k_{\mathrm{DSR}}<1$) represents a system whereby heat demand peaks during the electrical system peak. A DSR factor of unity ($k_{\mathrm{DSR}}=1$) represents a heat pump which effectively flattens heat demands over an entire day.

It is worth highlighting that non-daily metered (NDM) gas demand is estimated to be two-thirds composed of domestic space and water heating, with the remainder made up of small commercial or industrial users \cite{wilson2013historical}. We presuppose, however, that the underlying seasonality and temperature sensitivity of heating demand $H$ is much more likely to follow the NDM gas demand $G$ than the electrical demand $E$.

\subsection{Effects of Temperature on Demand}

It is common for SOs to account for the effects of weather on demand. In the case of GB, the electrical SO, National Grid ESO (NGESO), uses the Average Cold Spell (ACS) methodology, whilst the gas SO, National Grid Gas, uses the Composite Weather Variable (CWV). The former determines the correlation between daily temperature measurements and electrical demand using a linear regression based method \cite{ng2020acs}, whilst the latter uses a non-linear combination of weather and seasonal (i.e., temporal) parameters to derive an almost-linear fit between the CWV and gas demand \cite{xoserve2014autumn}.

In this work it is assumed that the demand $D$ is linearly dependent on temperature. By adding concurrent gas and electrical demand in \eqref{e:ted}, a least-squares linear fit is found of the form
\begin{equation}\label{e:t2d}
D = D_{0} + k_{T}T\,,
\end{equation}
where $k_{T}$ is the demand-temperature sensitivity coefficient, and $D_{0}$ represents temperature insensitive demand. This formulation is advantageous as it allows for long-term climate data to be incorporated (30 years of data are typically used to model the climate of a region \cite{ng2020acs}).

The form of the system margin $Z$ is found by combining \eqref{e:generation}, \eqref{e:findY} and \eqref{e:ted} along with the appropriate electrical heating demand model (\eqref{e:modelM0}, \eqref{e:modelM1} or \eqref{e:modelM2}) and temperature sensitivity model \eqref{e:t2d}. The system margin $Z$ is therefore given by
\begin{equation}\label{e:fullModel}
Z = X + Y - \left( D_{0} + k_{T}T\right)\,.
\end{equation}

\begin{figure}\centering
\subfloat[Gas-wind correlation]{\includegraphics[width=0.235\textwidth]{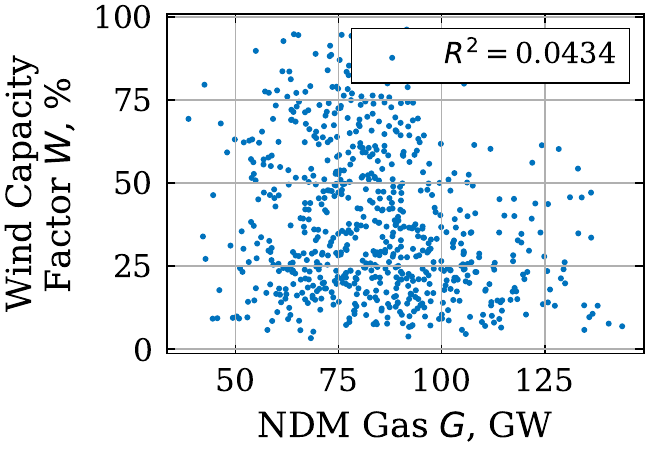}\label{f:pltResid6_G_W}}
~
\subfloat[Gas-temperature correlation]{\includegraphics[width=0.235\textwidth]{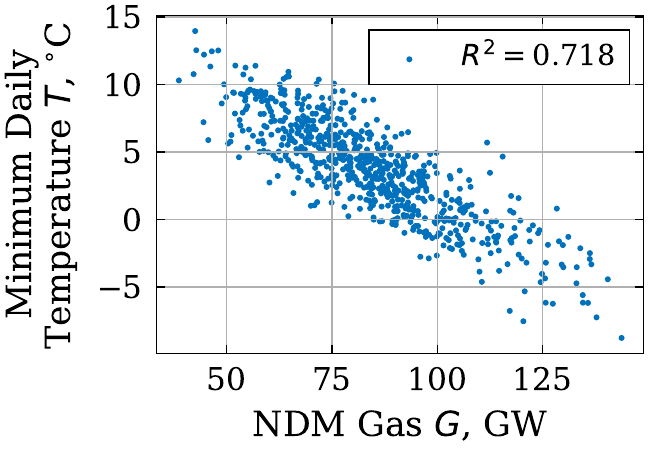}\label{f:pltResid6_G_T}}
\caption{Scatter plot of historic NDM gas demand $G$ (from \cite{ngg2020data,wilson2013historical}) against daily wind capacity factor $W$ (from \cite{staffell2016using}) and minimum daily temperature $T$ (from \cite{met2012met}) for five winters. The calculated coefficient of determination $R^{2}$ for the relation between variable pairs demonstrates that gas demand $G$ is much more weakly correlated with wind capacity factors $W$ than temperatures $T$.}
\label{f:pltResid6}
\end{figure}

\textit{Remark: Correlations Between Random Variables in \eqref{e:fullModel}.} The correlation between renewables (wind) $Y$ and electrical demand $E$ has been studied in detail in several recent works \cite{wilson2018use,bloomfield2018changing}. It is noted in those works that the correlation between wind and underlying demand is relatively weak. The CWV used by the UK gas industry does make use of a wind-chill factor \cite{xoserve2014autumn}. However, as with electrical demand $E$, historic NDM gas demand $G$ is much more weakly correlated with wind than temperature (see Fig. \ref{f:pltResid6}). Therefore, for the purposes of this exposition, only the correlation between temperature and demand is explicitly considered in \eqref{e:fullModel}, although future work could study these correlations in more detail. Given this independence, the PDF of $Z$ is determined by a convolution of the PDF of each of the random variables.

\subsection{System Adequacy Summary Statistics}

We shall consider the calculation of the peak demand, as well as two summary statistics to represent the likelihood and intensity of generation shortfall. The peak demand $P$ is found simply as the maximum demand over $N$ periods,
\begin{equation}\label{e:peakDemandDefn}
P = \max \{ D_{1},\,D_{2},\,\ldots,\,D_{N}\}\,,
\end{equation}
where each of the draws $D_{i}$ are assumed to be independent, with the CDF of this random variable denoted by $F_{P}$.

The first of the system adequacy metrics considered is the Loss of Load Expectation (LOLE), which is used to represent the likelihood of generation shortfall, and is typically reported in units of hrs/yr. This is calculated as
\begin{equation}\label{e:loleDefn}
\mathrm{LOLE} = \sum_{t=1}^{N} \mathbb{P}(Z_{t}<0) \,,
\end{equation}
where $Z_{t}$ represents independent draws of the system margin random variable $Z$, $N$ is the number of periods in a season, and $\mathbb{P(\alpha)}$ denotes the probability of some event $\alpha$.

The other calculated metric is the Expected Energy Unserved (EEU), which represents the intensity of generation shortfall, and is typically given in MWh/yr (or as a fraction of total system demand). This is found by calculating the expected value of the demand shortfall, given by
\begin{equation}\label{e:eeuDefn}
\mathrm{EEU} = \mathbb{E} \left( \sum_{t=1}^{N} \mathrm{max}\{0,-Z_{t}\} \right)\,,
\end{equation}
where $\mathbb{E}$ denotes the expectation operator, and $\mathrm{max}\{0,-Z_{i}\}$ is the shortfall at time period $t$.

\section{GB Demand, Generation and Scenario Data}\label{s:s3}

There are six random variables ($X,\,W_{\mathrm{on}},\,W_{\mathrm{off}},\,E,\,G,\,T$) that must be modelled for the calculation of the system margin $Z$ from \eqref{e:fullModel}, as well as the two system performance indices ($k_{\mathrm{DSR}},\,k_{\mathrm{COP}}$) and heat demand growth rate $\epsilon_{\Delta H}$. The sources for datasets for modelling are outlined in this section, with a summary of the data sources used for modelling random variables given in Table \ref{t:dataSources}. The peak season for which data are obtained is chosen to be the twenty weeks following the first Sunday of November \cite{sheehy2016impact}.

The electrical SO NGESO's Future Energy Scenarios (FES) \cite{ngeso2019fes} are used for modelling future scenarios. This details four credible pathways that NGESO envisions the GB energy system could follow, depending on the level of decentralisation and decarbonisation ambition. It describes four system pathways: `Two Degrees' (high decentralisation and decarbonisation), `Community Renewables' (low decentralisation, high decarbonisation),  `Consumer Evolution' (high decentralisation, low decarbonisation) and `Steady progression' (low decentralisation and decarbonisation).

\begin{table}
\centering
\caption{Summary of data sources for modelling random variables.}\label{t:dataSources}
\begin{tabular}{lll}
\toprule
Variable & Description & Reference \\
\midrule
$X$ & Conventional generation & \cite{ngeso2019fes,ng2013etsy,ofgem2014electricity,ofgem2020existing} \\
$W_{\mathrm{on}},\,W_{\mathrm{off}}$ & Hourly wind capacity factor & \cite{staffell2016using} \\
$E$ & Electrical demand & \cite{ngeso2020data}\\
$G$ & NDM gas demand & \cite{ngg2020data} \\
$T$ & Temperature & \cite{met2012met}\\
\bottomrule
\end{tabular}

\end{table}

\subsection{Rates of Heat Electrification and System COP}

To model credible rates of heat electrification, $\epsilon_{\Delta H}$, the rate of change of total commercial and residential (CR) gas demand is calculated across all four of the FES scenarios. We consider this value to be meaningful as it is assumed that the standard of living (and with it, underlying heating demand) is unlikely to change significantly. If there is a large increase in the efficiency of buildings when they are retrofitted with heat pumps, then this can be captured in the aggregate COP factor $k_{\mathrm{COP}}$.

The average rates of CR gas demand reduction of the two environmentally ambitious scenarios (`Two Degrees',`Community Renewables') scenarios are 9.5 TWh/yr and 9.9 TWh/yr (over the period to 2050). This is on a current NDM gas demand of $\epsilon_{G} = 440$ TWh/yr, and a GB electrical transmission system demand of $\epsilon_{E} = 285$ TWh/yr. It is worth noting that the rates of reduction vary in CR gas demand across these time periods, with a maximum rate of reduction of 18.7 TWh/yr, which occurs in the `Two Degrees' scenario, and a smallest rate -0.97 TWh/yr in the `Steady Progression' scenario. Based on these numbers, we consider an increase in electrified heating demand of $\epsilon_{\Delta H} = 12.5$ TWh/yr each year. This is intended to represent a concerted effort to electrify current CR heat demand.

The aggregate coefficient of performance parameter $k_{\mathrm{COP}}$ is typically between 1.5 and 4.0 for individual systems, depending the technology and external temperature \cite{staffell2012review}. In this work we assume an aggregate winter COP of $k_{\mathrm{COP}}=1.9$. Although aggregate heat pump demand could show peaks of 200\% of daily demand \cite{clegg2019integrated}, a more optimistic DSR coefficient of $k_{\mathrm{DSR}}=1.0$ is chosen as a central estimate.

\subsection{Demand and Temperature Data}
\begin{figure}\centering
\includegraphics[width=0.44\textwidth]{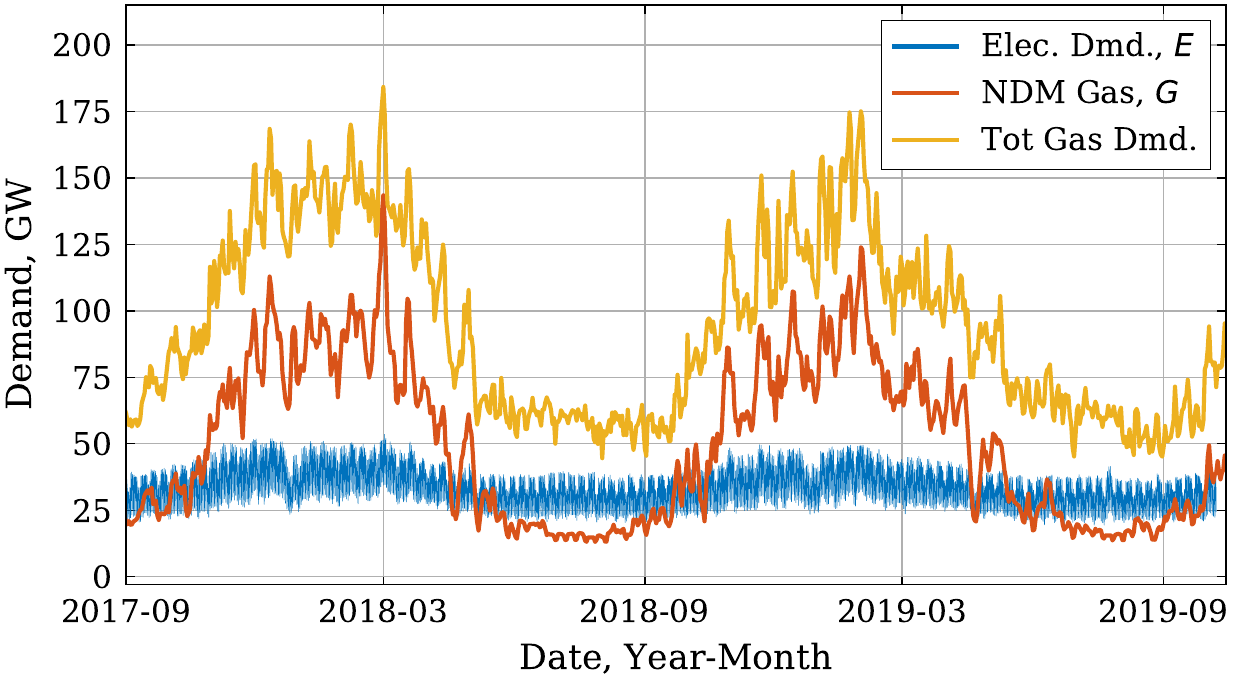}
\caption{Historic electrical demand $E$ and NDM gas demand $G$, demonstrating the latter has (i) a larger yearly energy demand, (ii) a significantly increased peak demand, and (iii) a stronger seasonality than electrical demand. Also plotted is the total GB transmission system gas demand, demonstrating that much of the seasonality of gas demand is as a result of the NDM component.}\label{f:riskdayFig1}
\end{figure}

Five years of concurrent historic demand data are available from the GB gas and electrical SOs \cite{ngeso2020data,ngg2020data}. To determine underlying electrical demands, least-squares linear regression is first used to remove the long-term trend, after which the resulting demand is scaled so that the peak matches industry estimates of the `true' peak (from \cite{ngeso2019ecr}). As noted in \cite{wilson2013historical}, both the size and variability of NDM gas demand is much greater than electrical demand (see Fig. \ref{f:riskdayFig1}). Additionally, weather events impact on gas demand in a much more pronounced way (such as the `Beast from the East' cold snap, visible as a sharp increase in gas demand at the start of March 2018). 

To obtain daily temperature values $T$, the data from nationally approved weather stations are obtained from the MIDAS database \cite{met2012met} (the list of approved stations is publicly available from \cite{elexon2019approved}). Where there is missing data from the primary weather station, data from the approved secondary station is usually used. The fourteen zoned temperature measurements are converted to a single temperature variable by taking a weighted sum, with each weight scaled to be proportional to the total zonal peak demand from \cite{ngeso2019fes}.
\subsection{Generation Data}

The breakdown of future conventional and renewable generation mixes are taken from the FES Five Year Forecast. Forced outage rates for conventional generators are taken from OFGEM data \cite[Table 1]{ofgem2014electricity}, with availabilities between 81\% and 97\% depending on the technology. The availability of interconnectors is estimated to be 80\%, following interconnector availability requirements for the GB cap-and-floor regime \cite{ofgem2016cap}. Estimated offshore and onshore wind capacity factors are taken from \cite{staffell2016using}, which are then combined with historic or predicted onshore/offshore wind capacities to determine the PDF describing the renewable generation $Y$.

The current edition of FES is for 2019, and so 2018 represents the base year, with five year forecasts running to 2023. As in \cite{sheehy2016impact}, the size and number of individual generating units at present are determined from National Grid's 2013 high decarbonisation/high prosperity `Gone Green' scenario for the 2013/2014 season \cite{ng2013etsy}\footnote{More recent editions of FES only consider power stations, rather than individual generator units; note that the data used are not publicly available due to the politically sensitive nature of system adequacy calculations.}. The distribution of generating units are found by randomly assigning individual units from the historic 2013 dataset, until the generation quota matches the generating fleet totals from NGESO's Five Year Forecast \cite{ngeso2019fes}. Future interconnector capacities are taken from \cite{ofgem2020existing}. 

\section{Results}\label{s:s4}

The primary aim of this work is to study \textit{relative} rather than \textit{absolute} system adequacy measures, as policy decisions required to drive heat electrification should improve the prospects of generators in a market-based system, leading to an increase in supply. We consider first LOLE and EEU metrics, with the aim of demonstrating that the underlying assumptions on the shape of demand make a significant impact on calculated system risk indices. Secondly, we study the sensitivity of the models to temperature, to evaluate the vulnerability of the system to prolonged cold snaps that can occur across Western Europe. In each case, the base model $\mathcal{M}_{0}$ is compared against the electricity-scaled model $ \mathcal{M}_{1} $ and the gas-scaled model $ \mathcal{M}_{2} $.

\subsection{System Adequacy Measures}

\begin{figure}\centering
\subfloat[Calculated LOLE]{\includegraphics[width=0.22\textwidth]{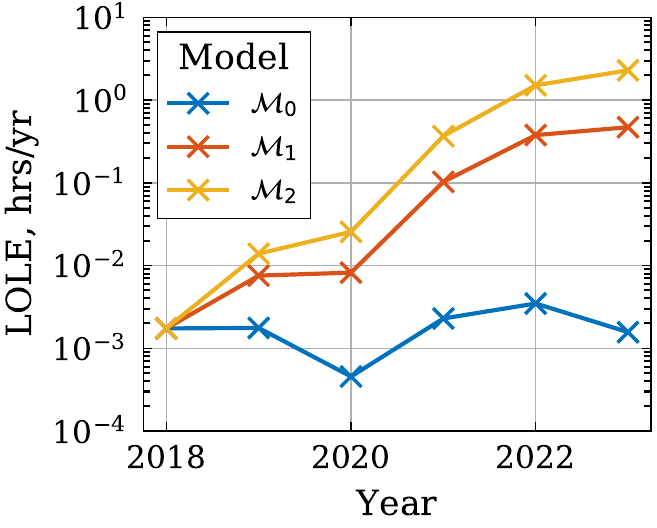}\label{f:plt_xAll_lole}}
~
\subfloat[Calculated EEU]{\includegraphics[width=0.215\textwidth]{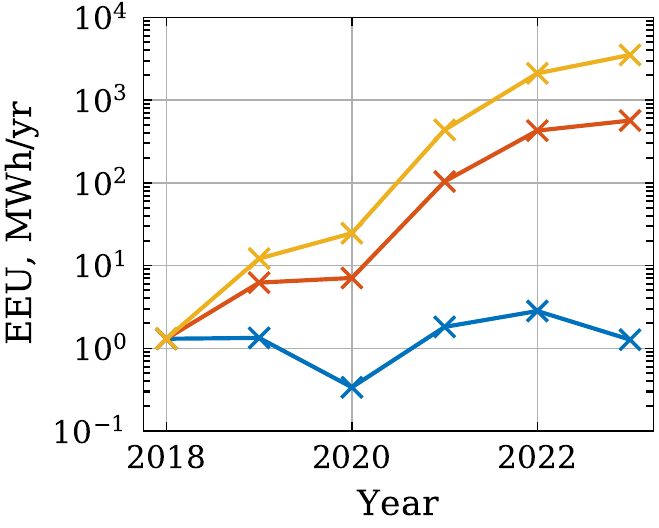}\label{f:plt_xAll_eeu}}
\caption{Calculated values of the LOLE \eqref{e:loleDefn} and the EEU \eqref{e:eeuDefn}, across years (with DSR coefficient, $k_{\mathrm{DSR}}=1.0$, system COP $k_{\mathrm{COP}}=1.9$) and heat demand models $\mathcal{M}_{0}$-$\mathcal{M}_{2}$.}
\label{f:plt_xAll}
\end{figure}

The LOLE of the system, as described in Section \ref{s:s3}, is plotted in Fig. \ref{f:plt_xAll_lole}. The base case (model $ \mathcal{M}_{0} $) shows a consistently low LOLE, well below the GB LOLE standard of 3 hrs/yr \cite{decc2013reliability}. With this model, it is only changes to the Total Generation \eqref{e:generation} which drive changes in this value, as both the electrical and heating demand remain constant. Although the forecast of the conventional generation capacity $X$ predicts the retirement of several large thermal generators, this is offset by the increased interconnector capacity (increasing from 4.0 GW in 2018 to 11.7 GW in 2023) and increased renewable generation $Y$ (total wind capacity increases from 20.9 GW to 27.4 GW).

Both the electrical-scaled and gas-scaled models $ \mathcal{M}_{1},\,\mathcal{M}_{2} $ show an increase in the LOLE, although the rate of increase in the gas-scaled model $\mathcal{M}_{2}$ is much faster than the electrical-scaled model $ \mathcal{M}_{1} $. By 2023, the LOLE is four times greater for the gas-scaled model $ \mathcal{M}_{2} $. The EEU shows a similar story, as shown in Fig. \ref{f:plt_xAll_eeu}. There is a consistently small EEU for the base model $ \mathcal{M}_{0} $, but the gas-scaled model $ \mathcal{M}_{2} $ shows significantly increased EEU compared to the electricity-scaled model $ \mathcal{M}_{1} $.

\begin{figure}\centering
\includegraphics[width=0.42\textwidth]{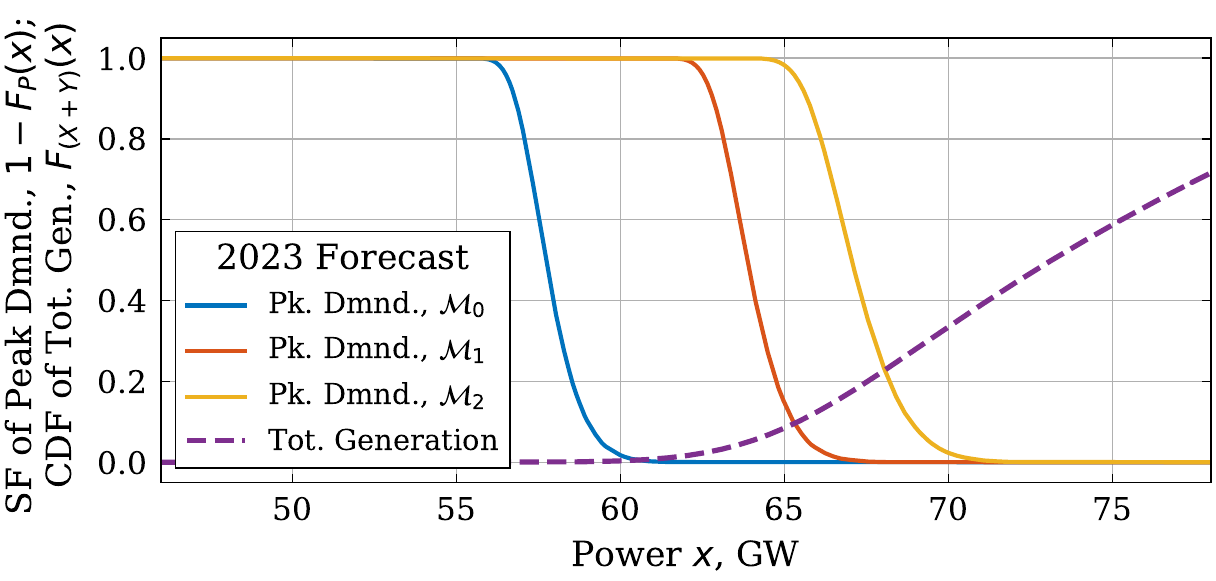}
\caption{The survival function (SF) of the peak annual demand $P$, $1-F_{P}(x)$ and the CDF of the total generation $X+Y$, $F_{(X+Y)}$ for the 2023 forecast.}\label{f:genPkLds}
\end{figure}

In Fig. \ref{f:genPkLds} the survival function (1-$F_{P}$) of each of the model's system peak demands $P$ \eqref{e:peakDemandDefn} are plotted against the CDF of the Total Generation \eqref{e:generation}, for the 2023 forecast. From this figure, it is immediately apparent why the system margin $Z$ results in much larger LOLE and EEU values for the gas-scaled model $ \mathcal{M}_{2} $. The 95\% quantile of the annual peak $P$ (i.e., the 1-in-20 peak) increases from 59.4 GW in the base case (model $\mathcal{M}_{0}$) to 65.7 GW using model $\mathcal{M}_{1}$, or to 69.4 GW using model $\mathcal{M}_{2}$. Additionally, the spread of peak demands has increased: the interquartile range of the peak demand ranges from 1.16 GW for the base model $ \mathcal{M}_{0} $ to 1.29 GW in the electricity-scaled model $ \mathcal{M}_{1} $, whilst it is 1.67 GW in the gas scaled model $ \mathcal{M}_{2} $. It is worth re-emphasizing that the underlying total energy modelled by the gas-scaled and electric-scaled models is identical, and it is only the difference in the heat demand profile $H$ that leads to the observed changes.

\subsection{Sensitivity of Demand to Temperature}

\begin{figure}\centering
\includegraphics[width=0.42\textwidth]{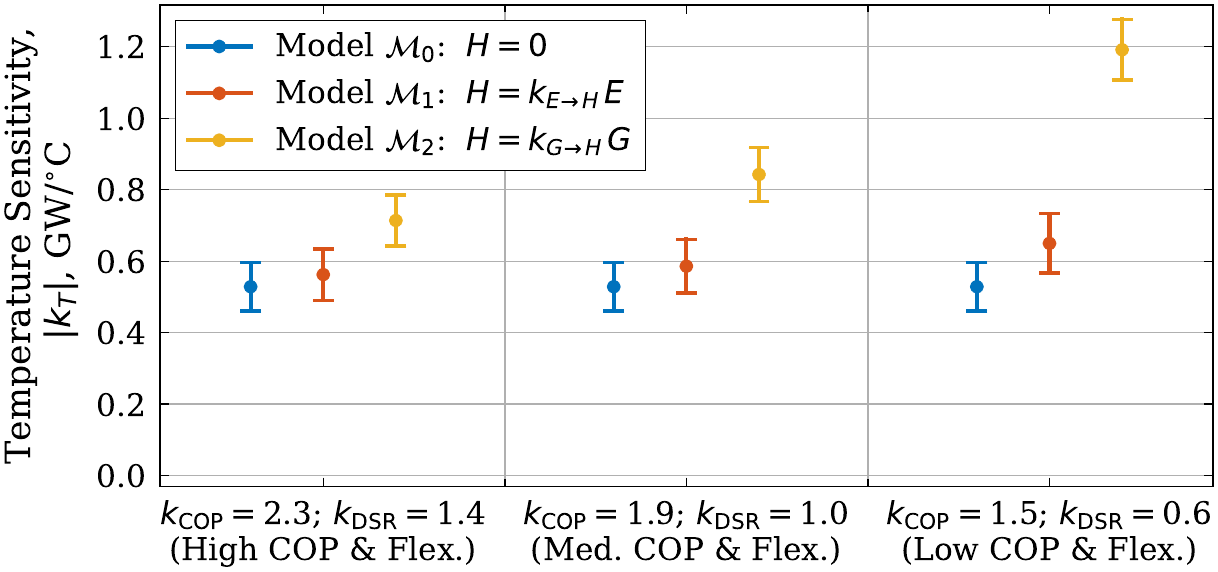}
\caption{Comparing demand-temperature sensitivity $k_{T}$, following five years of heat demand growth (i.e. in 2023), across system models and parameters. The whiskers represent the 95\% confidence intervals ($\pm2$ times the standard error of the fit to $k_{T}$). All values of $k_{T}$ are negative (i.e., demand increases as temperature decreases).}\label{f:k_T_fig}
\end{figure}

Finally, we consider the calculation of the calculation of temperature sensitivity factors for each of the models, plotted in Fig. \ref{f:k_T_fig}, for the final year of the forecast (2023). As well as the values of the DSR and COP coefficients $k_{\mathrm{DSR}},\,k_{\mathrm{COP}}$ presented in the prior results (the `medium COP and flexibility' values), two additional sets of coefficients are chosen to represent a range of potential future system parameters.

The figure shows the base model $\mathcal{M}_{0}$ maintaining a temperature sensitivity close to 0.5 GW/$^{\circ}$C, as the underlying composition of the demand does not change. The increase in temperature sensitivity from the electricity-scaled model $ \mathcal{M}_{1} $ to the gas-scaled model $ \mathcal{M}_{2} $ is stark. In the low COP/flexibility model, representing an inefficient system, the thermal sensitivity reaches 1.2 GW/$^{\circ}$C, more than doubling the demand-temperature sensitivity. Referring back to historic temperatures from Fig. \ref{f:pltResid6_G_T}, this could result in 24 GW swings in demand that are entirely weather dependent.

\section{Conclusions}\label{s:s5}

This work has studied the impact of the electrification of heat on system adequacy calculations. It has been demonstrated that assumptions about the system demand profile representing increased heat demand lead to very different calculations of risk, in terms of LOLE and EEU, as well as in calculations of sensitivities of the system to weather events. For example, 1-in-20 peak demand increases 60\% faster when a gas rather than electricity profile is assumed, whilst a low system efficiency scenario shows a doubling of temperature sensitivity in just five years. The analysis makes very clear the huge capacity value that gas provides to the GB heat sector, and demonstrates the challenges of large-scale electrification of heat, even at relatively small penetration levels. 

Although the presented method is relatively simple, it represents a link between data-driven system adequacy methods and traditional generative heat demand modelling. Parameters modelling the system COP and impacts of heat DSR allow for sensitivity analysis of results to be computed efficiently. Future work could combine the methods presented here with more detailed physical models--heat demand rebound effects could result in more serious and prolonged loss-of-load incidents, whilst solar, energy storage and electrical DSR will can also all impact on adequacy. The time-coupled nature of these technologies suggests that a time-sequential model could be used to study how to mitigate these risks, whilst also providing a more accurate calculation of system adequacy metrics.

\bibliographystyle{IEEEtran}
\bibliography{refs}

\end{document}